\documentclass[%
prfluids,%
 longbibliography,
 amsmath,amssymb,
preprint,%
floatfix]{revtex4-1}
\usepackage{color}
\usepackage{natbib}
\usepackage{graphicx}% Include figure files
\usepackage{dcolumn}% Align table columns on decimal pointf
\usepackage{bm}% bold math
\preprint{AIP/123-QED}
\newcommand\Ra{\mbox{\textit{Ra}}}  % Rayleigh number
\newcommand\Pran{\mbox{\textit{Pr}}}  % Prandtl number
\newcommand\Nu{\mbox{\textit{Nu}}}  % Nusselt number

\begin{document}

\preprint{APS/123-QED}

\title{Subcritical convection in an internally heated layer}

\author{Linyan Xiang}
\author{Oleg Zikanov} \email{zikanov@umich.edu}
\affiliation{Department of Mechanical Engineering, University of Michigan-Dearborn, 48128-1491 MI, USA}

\date{\today}

\begin{abstract}
Thermal convection in a horizontal layer with uniform internal heating and stress-free constant-temperature boundaries is analyzed numerically. The work is  motivated by the questions arising in development of liquid metal batteries, in which convection is induced by the Joule heating of electrolyte. It is demonstrated that three-dimensional convection cells exist at subcritical Rayleigh numbers.
\end{abstract}


\maketitle

\section{Introduction}\label{sec:intro}
We consider the thermal convection caused by uniform volumetric heating in a horizontal layer of a fluid. The work is motivated by the current efforts to develop and commercialize liquid metal batteries - devices for short-term large-scale storage of electric energy \cite{Kim:2013}. One often considered version of the battery can be schematically  viewed as a vertically stably stratified system of three horizontal liquid layers: a layer of a heavy metal (e.g. Bi, Sb or PbSb) at the bottom, a layer of a light metal (e.g. Na, Li or Mg) at the top, and a thin layer of a molten-salt electrolyte of intermediate density sandwiched in the middle. Strong (up to or even exceeding 1 A/cm$^2$) nearly vertical steady electric current passes through the system during charging and discharging.

It has been demonstrated in the recent numerical simulations \cite{Shen:2016,Koellner:2017}  that the internal Joule heating of the poorly electrically conducting electrolyte causes thermal convection flow in this layer. Penetrative convection is also induced in the top and, with a much smaller amplitude, bottom metal layers. The simulations have shown that the flow exists in small laboratory prototypes and should become strong and turbulent in large commercial batteries. It has also been shown in  \cite{Shen:2016} that, although a  magnetic field is always induced within the battery by the electric currents flowing through it and through the adjacent connectors, the resulting magnetohydrodynamic effect on the thermal convection flow is negligibly weak unless batteries of very large size are considered. 

In this work we explore the phenomenon using an idealized model of a single layer of electrolyte. The effect of the top and bottom metal layers is approximated by stress-free non-deformable horizontal boundaries. This is a reasonably good approximation of reality because, as discussed in \cite{Shen:2016}, the strong stable density stratification between the layers implies that the perturbation of the interface caused by the convection flow is insignificant except in very larger batteries with strong turbulent convection.
Furthermore, as an approximation justified by much higher thermal conductivity and thermal capacity of the metals, the boundaries are assumed to be maintained at a constant temperature. 

The natural thermal convection in horizontal layers with internal heating has been studied before (see, e.g., classical experiments  \cite{Kulacki:1972}, computations \cite{Goluskin:2015}, and  review  \cite{GoluskinBook:2015}). Still, unanswered questions remain. In particular, the subcritical behavior of  the system considered in this paper is unknown. It has been shown using the spectral collocation method that the pure conduction state becomes linearly unstable at $\Ra_L=16992.2$ (the Rayleigh number is defined in section \ref{sec:model}), while the energy stability analysis of the same state gives $\Ra_E=10618.1$ (see \cite{GoluskinBook:2015} and references therein). The inequality $\Ra_E<\Ra_L$, which is also found in other systems with internal heating  \cite{GoluskinBook:2015}, means that convection flows may exist as subcritical regimes at $\Ra_E<\Ra<\Ra_L$. The existence was computationally confirmed for the case of no-slip walls and infinite Prandtl number \cite{Tveitereid:1978}. No such confirmations and no descriptions of the subcritical states are, however, available for many other configurations, in particular, for the configuration considered here.

Our interest in the flow states at small $\Ra$ is related to the operation of liquid metal batteries. It is anticipated  that the electrolyte layer will be thin, with the thickness of the order of few mm, in order to minimize the Joule heat losses \cite{Kim:2013}. We can estimate the Rayleigh number using, as an example, the physical properties of LiCl-KCl at 450$^{\circ}$ C (see \cite{Shen:2016}). In this case, $\Ra_L=16992.2$ corresponds to $d\approx 3.6$ mm, a realistic value.

\section{Physical model and numerical method}\label{sec:model}
The Boussinesq approximation is applied. The typical scales used to derive the non-dimensional governing equations are $\Delta T\equiv q d^2\kappa^{-1}$ for temperature, $U\equiv \sqrt{\alpha g \Delta T d}$ for velocity, $d$ for length, $d/U$ for time, and $\rho U^2$ for pressure. Here, $d$ is the height of the layer, $q=const$ is the internal heating rate per unit volume, and $\kappa$, $\alpha$, $\rho$,  $\nu$, and $\chi$ are the physical coefficients of thermal conductivity, thermal expansivity, density, kinematic viscosity, and thermal diffusivity, all assumed constant. 

The equations and boundary conditions are:
\begin{eqnarray}
\label{eq1}
 & & \frac{\partial \bm{u}}{\partial t} +
(\bm{u}\cdot \nabla)\bm{u} = -\nabla p
+  \sqrt{\frac{\Pran}{\Ra}} \nabla^2 \bm{u} + T\bm{e}_z
,\\
\label{eq2}
& & \nabla \cdot\bm{u}  =  0,\\
\label{eq3} 
& & \frac{\partial T}{\partial t} +
\bm{u}\cdot \nabla T = \frac{1}{\sqrt{\Ra\Pran}}\left(\nabla^2 T + 1\right),\\
\label{eq4}
& & w=\frac{\partial u}{\partial z}=\frac{\partial v}{\partial z}=T=0 \;\; \mbox{at} \; z=\pm \frac{1}{2}.
\end{eqnarray}
Here $\bm{u}=(u,v,w)$, $p$, and $T$ are the fields of velocity, pressure and deviation of temperature from its boundary value. 
The $z$-coordinate is in the upward vertical direction.  The non-dimensional flow domain is a cuboid $-1/2\le z\le 1/2$, $0\le x,y\le L$.
The boundaries at $x=0,L$ and $y=0,L$ are subject to periodicity conditions. 

The non-dimensional parameters are the Rayleigh and Prandtl numbers
\begin{equation}
\label{param}
\Ra\equiv \frac{g\alpha d^3\Delta T}{\chi \nu}, \;\; \Pran\equiv \frac{\nu}{\chi}.
 \end{equation}
The constant value $\Pran=3.3$ corresponding to LiCl-KCl at about 450$^{\circ}$C  is used.

The pure conduction state, stability of which is analyzed, is
\begin{equation}
\label{base}
\bm{u}_0=0, \:\: T_0=\frac{1}{8}-\frac{1}{2}z^2.
\end{equation}
In the following discussion, the computed convection flows are quantified using the average kinetic energy, vertical heat flux, and Nusselt number:
\begin{equation}
\label{quant}
E=\langle u^2+v^2+w^2\rangle, \: W=\langle wT\rangle, \: \Nu=\frac{1}{12\langle T\rangle},
\end{equation}
where $\langle \ldots \rangle$ stands for volume averaging over the entire computational domain. Time averaging over long periods of evolution of a fully developed flow is also applied for unsteady regimes. Integration of the heat equation shows that the average heat fluxes through the top and bottom boundaries are
\begin{equation}
\label{fluxes}
Q_{\mbox{top}}=\frac{1}{2}+W, \: Q_{\mbox{bottom}}=\frac{1}{2}-W.
\end{equation}
  
The problem (\ref{eq1})-(\ref{eq4}) is solved numerically in three-dimensional and two-dimensional (with the fields of solution uniform in one of the horizontal directions) formulations. The method is based on the conservative, structured-grid,  finite-difference scheme of the second order in space and time introduced as the Scheme B in \cite{Krasnov:FD:2011}. Detailed descriptions can be found in this reference and in later works, in which the scheme is applied to incompressible flows with shear, thermal convection, and magnetohydrodynamic effects (see, e.g. \cite{Krasnov:2012,Zhao:2012,Zhang:2014,Lv:2014}). 

The results reported in the next section are computed in the domains with $L=\lambda_L$ (the small domain) and $L=4\lambda_L$ (the large domain), where $\lambda_L=2.075$ is the wavelength of the neutral perturbations of the linear stability problem at $\Ra=\Ra_L$ \cite{GoluskinBook:2015}. The computational grids are uniform and have
$N_z=32$ points in the vertical direction and $N_{\bot}=64L/\lambda_L$ points in the horizontal directions. The reason for such relatively crude grids is the  long intervals of time $t$ needed in the simulations to obtain the subcritical flow regimes. {We have performed a grid sensitivity study and found that increasing the resolution to $N_z=64$, $N_{\bot}=128L/\lambda_L$  does not lead to any significant changes of the flow's characteristics. Decreasing it to $N_z=16$, $N_{\bot}=32L/\lambda_L$  does not change the qualitative behavior and changes the quantitative characteristics $E$, $W$, and $\Nu$ by a few percent. Furthermore, we used the numerical model to compute the critical Rayleigh number of the linear instability $\Ra_L$ for the small two-dimensional domain. For this purpose, the evolution of the small-amplitude random perturbations added to the base flow was computed. The growth or decay rate estimated on the basis of the kinetic energy curves $E(t)$ were used to determine $\Ra_L$.  The results were 16952 at $N_z=32$, $N_{\bot}=64L/\lambda_L$, 16983 at $N_z=64$, $N_{\bot}=128L/\lambda_L$, and 16881 at $N_z=16$, $N_{\bot}=32L/\lambda_L$, which can be compared with the  value   $16992.2$ obtained in the eigenvalue analysis based on the spectral collocation discretization \cite{GoluskinBook:2015}. In the following discussion, the value $\Ra_L=16952$ obtained at the same level of resolution as in the majority of the reported simulations is used as an approximation of the linear stability threshold. }

\section{Results and discussion}\label{sec:results}
The simulations were performed in the way of solving the full system of unsteady governing equations (\ref{eq1})-(\ref{eq4}). In a few runs carried out at $\Ra>\Ra_L$ we reproduced the linear instability and the subsequent transition by adding small-amplitude random perturbations to the base state and computing the solution till saturation to a stable supercritical state. The majority of runs were performed to find other nonlinear convection solutions in the continuation manner, i.e. by starting with an already computed solution, decreasing $\Ra$ by a small increment, and calculating the flow's evolution until an approximation of a new state is achieved. The smallest increment $\Delta \Ra=3$ was used at the approach to the limit points of the solution branches. The flow's adjustment after the change of $\Ra$ was, typically, slow, requiring many thousands of time units. For this reason, in the computationally more demanding large-domain simulations, only some of the flow states, including the states near the limit point, were calculated {to full convergence}. For the other states, the approximations obtained after 3000 time units of evolution were accepted. 

\begin{table}
\renewcommand{\arraystretch}{0.8}
  \centering
  \begin{tabular}{cc|ccc}
     $\Ra$ & $\Ra-\Ra_L$ & $E$ & $W$ & $\Nu$  \\
     \hline
  & & \multicolumn{3} {l} {Small two-dimensional domain}\\
19800 & 2848	 & $2.15\times 10^{-4}$  & $6.68\times 10^{-5}$	& 1.050\\	  
17820 & 868	 & $8.14\times 10^{-5}$  & $2.59\times 10^{-5}$	& 1.019 \\	
17160  & 208	 & $2.17\times 10^{-5}$  & $6.95\times 10^{-6}$	& 1.005	\\
16995 & 43	 & $4.10\times 10^{-6}$  & $1.31\times 10^{-6}$ & 1.002\\
16955  &  3 & $4.14\times 10^{-7}$  & $1.31\times 10^{-7}$ & 1.001 \\
     \hline
 & & \multicolumn{3} {l} {Small three-dimensional domain}\\
 19800 & 2848	& $2.91\times 10^{-5}$ &  $8.90\times 10^{-5}$ &  1.052 \\
17820 & 868 &	$1.87\times 10^{-5}$  & $5.97\times 10^{-5}$  & 1.035 \\
17160 & 208 &	$1.38\times 10^{-4}$  & $4.51\times 10^{-5}$  & 1.027 \\
16995 & 43 &	$1.24\times 10^{-4}$  & $4.05\times 10^{-5}$  & 1.025 \\
16830 & -122 &	$1.08\times 10^{-4}$  & $3.54\times 10^{-5}$  & 1.022 \\
16665 & -287 & 	$8.83\times 10^{-5}$ &  $2.92\times 10^{-5}$  & 1.018 \\
16533 & -419 & $6.70\times 10^{-5}$  & $2.22\times 10^{-5}$  & 1.015 \\
16500 & -452 & $5.90\times 10^{-5}$  & $1.96\times 10^{-5}$  & 1.013 \\
16484 & -468 &	$5.34\times 10^{-5}$ & $1.78\times 10^{-5}$  & 1.012 \\
16470 &   -482 &  $4.44\times 10^{-5}$  & $1.48\times 10^{-5}$  & 1.010\\
     \hline
  & & \multicolumn{3} {l} {Large three-dimensional domain}\\
17820 & 868 & $	\underline{1.80\times 10^{-4}}$ &	$\underline{5.59\times 10^{-5}}$ &	\underline{1.032} \\	
17160 & 208 &	 $\underline{1.40\times 10^{-4}}$ &  $\underline{4.41\times 10^{-5}}$ &	\underline{1.025} \\	
16665 & -287 & $1.08\times 10^{-4}$ & $3.35\times 10^{-5}$ & 1.019 \\
16517 & -435 & 	$9.89\times 10^{-5}$ &	$3.08\times 10^{-5}$ &	1.017 \\	
16302 & -650 & 	$8.21\times 10^{-5}$ &	$2.59\times 10^{-5}$ &	1.015 \\	
16170 & -782 & 	$6.42\times 10^{-5}$ &	$2.00\times 10^{-5}$ &	1.012 \\
16015 &  -937 &   $3.88\times 10^{-5}$ &   $1.18\times 10^{-5}$ &   1.007\\
15936 & -1016 & $8.15\times 10^{-6}$ & $2.45\times 10^{-6}$ & 1.0023 \\
15932 & -1020 &$7.97\times 10^{-6}$ & $2.42\times 10^{-6}$ & 1.00225\\
 15926 & -1026 & $7.40\times 10^{-6}$ & $2.25\times 10^{-6}$ & 1.0022
\end{tabular}
  \caption{Kinetic energy $E$, vertical heat flux $W$ and the Nusselt number $\Nu$ (see (\ref{quant})) of computed stable convection flows. All fully accurately computed data are shown for the three-dimensional large domain. Selected data are shown for the three- and two-dimensional small domains. The data are obtained by volume averaging.  In the case of unsteady solutions indicated by the underline, time-averaging over not less than 10000 time units is also applied. The value $\Ra_L=16952$ found computationally with the same numerical resolution is used to calculate sub- and super-criticality.}
  \label{tab1}
\end{table}

\begin{figure}
\begin{center}
\includegraphics[width=0.7\textwidth]{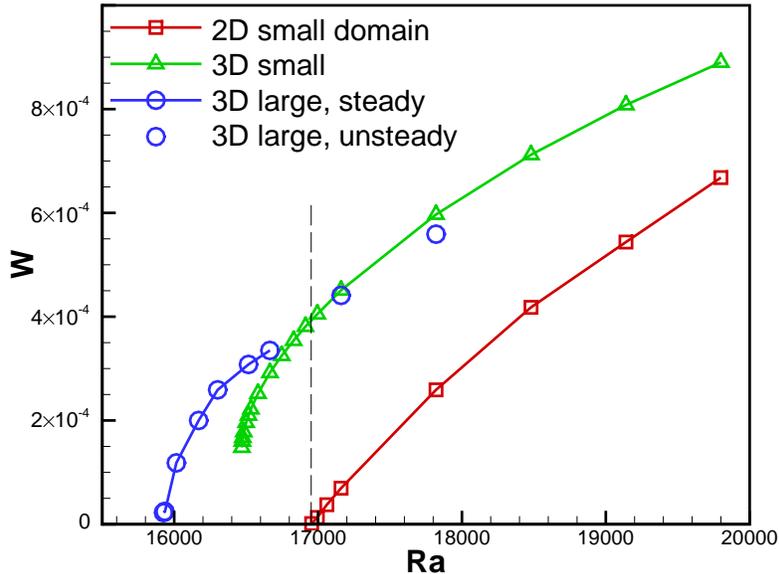}
\caption{Volume- and, for unsteady regimes, time-averaged vertical heat flux $W$ (see (\ref{quant})) of computed stable convection solutions.  The dashed line indicates the linear instability threshold $\Ra_L=16952$ found computationally with the same numerical resolution.}
\label{fig1}
\end{center}
\end{figure}

The main results of the simulations are summarized in Fig.~\ref{fig1} and Table \ref{tab1}. Integral characteristics (\ref{quant}) of the stable convection solutions are shown. They are obtained by volume averaging. Additional time-averaging   over not less than 10,000 time units is applied for the unsteady solutions found in the large domain at $\Ra > 16665=\Ra_L-287$. 

Only one branch of stable convection solutions is found for each computational domain. At $\Ra>\Ra_L$, identical solutions of the branch are found by both the methods: starting with a slightly perturbed base state, and continuing from a solution at a close $\Ra$. 
In three-dimensional simulations, the branch extends into the subcritical region $\Ra<\Ra_L$. No continuation into the subcritical region is found in the two-dimensional simulations. 

{It should be stressed that our method of analysis  allows us to compute only stable solutions. For this reason, the subcritical branches can only be followed into the subcritical zones till the limit points, which are found as $\Ra_{lim}=16470=\Ra_L-482$ for the small domain and $\Ra_{lim}=15926=\Ra_L-1026$ for the large domain. Decrease of $\Ra$ below these limits results in the rapid decay of convection flow structures and convergence to the base state (\ref{base}). The implied unstable parts of the convection solution branches connecting the limit points with the bifurcation point at $\Ra=\Ra_L$ cannot be followed.}

\begin{figure}
\begin{center}
\includegraphics[width=0.6\textwidth]{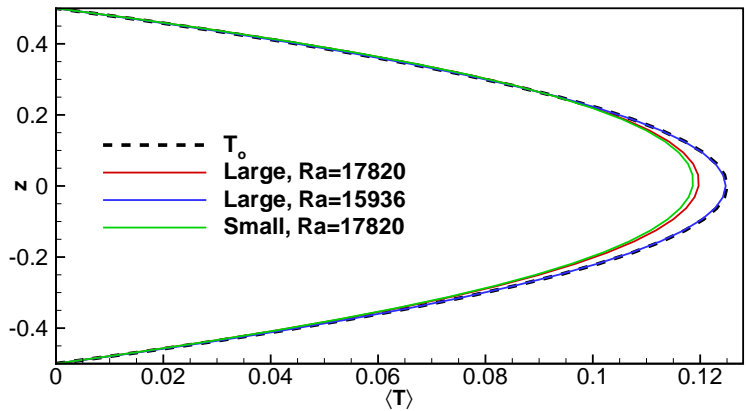}
\caption{{Vertical profiles of the horizontally averaged temperature in computed three-dimensional convection flows at supercritical $\Ra=17820$ (for large and small domains) and subcritical $\Ra=15936$ (for large domain). The large domain curve at  $\Ra=17820$ is obtained as a result of time averaging. The parabolic base temperature profile $T_0(z)$ (see (\ref{base})) is shown for comparison.}}
\label{fig2}
\end{center}
\end{figure}

Table \ref{tab1} shows that the kinetic energy and the generated additional heat flux are low, with the Nusselt number remaining close to 1. {The horizontally averaged profiles of temperature shown in Fig.~\ref{fig2} are close to the base profile (\ref{base}). We conclude that in the parameter range considered here the convection flows are weak.}

\begin{figure*}
\begin{center}
\includegraphics[width=0.35\textwidth]{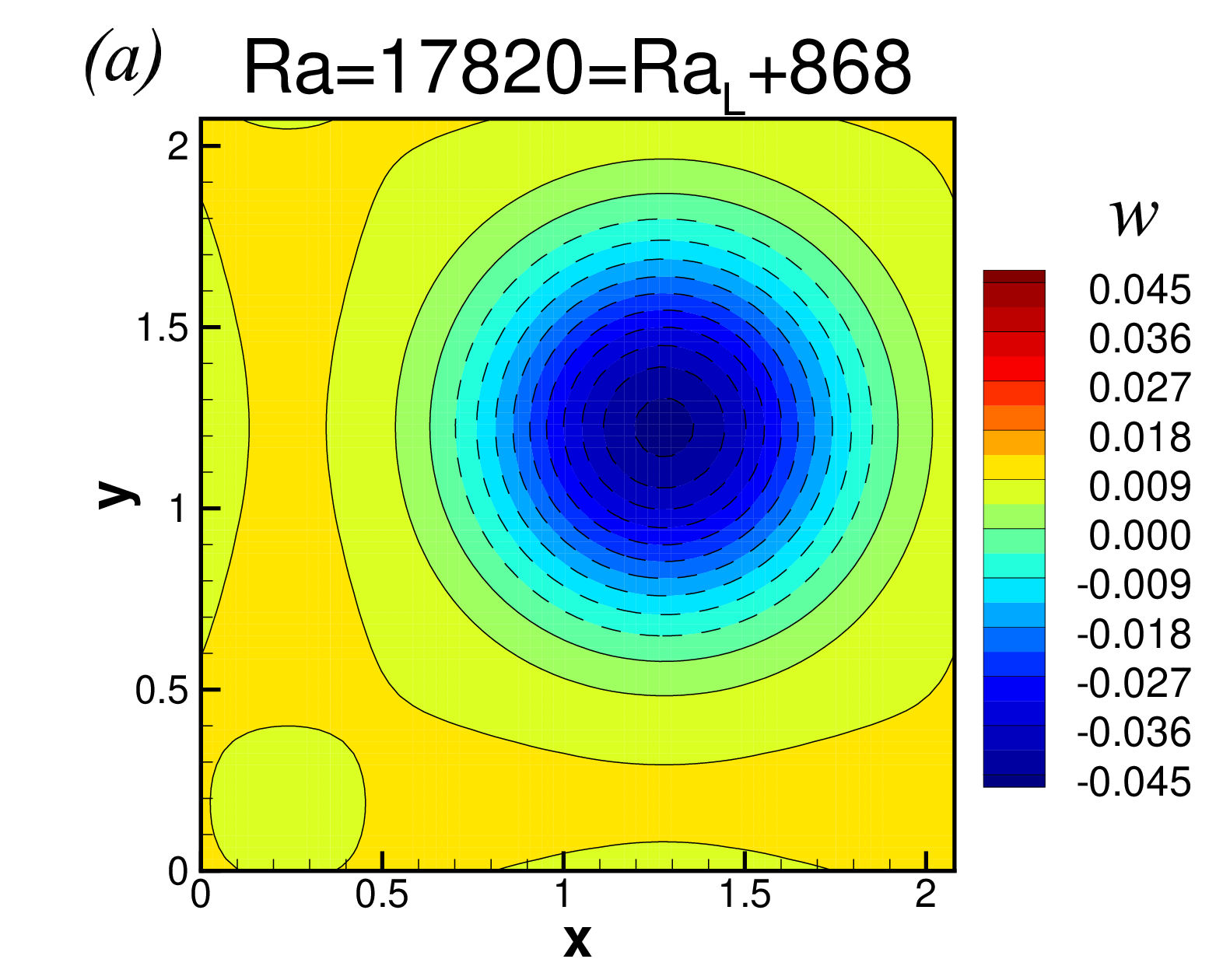}
\includegraphics[width=0.63\textwidth]{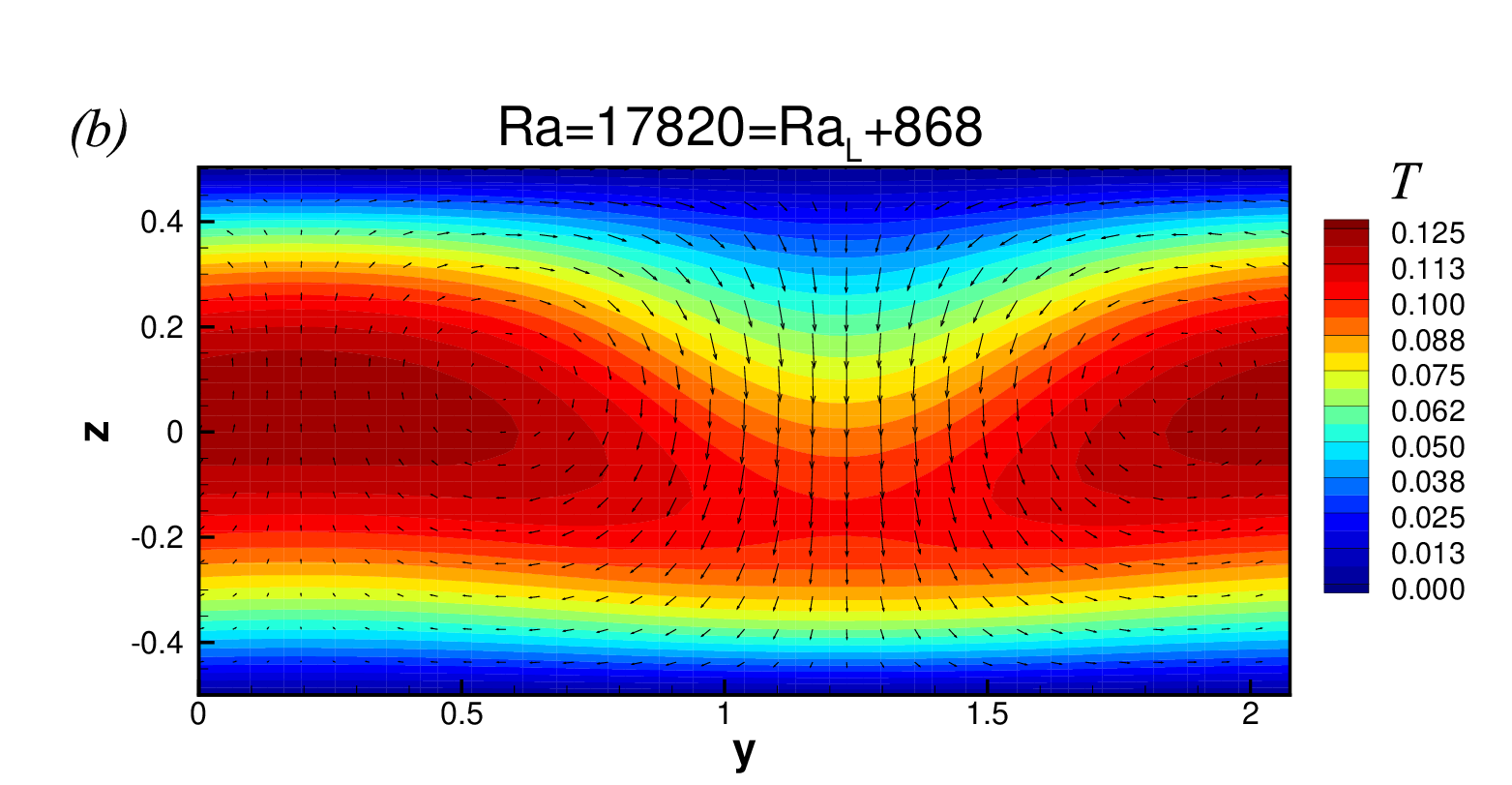}\\
\includegraphics[width=0.49\textwidth]{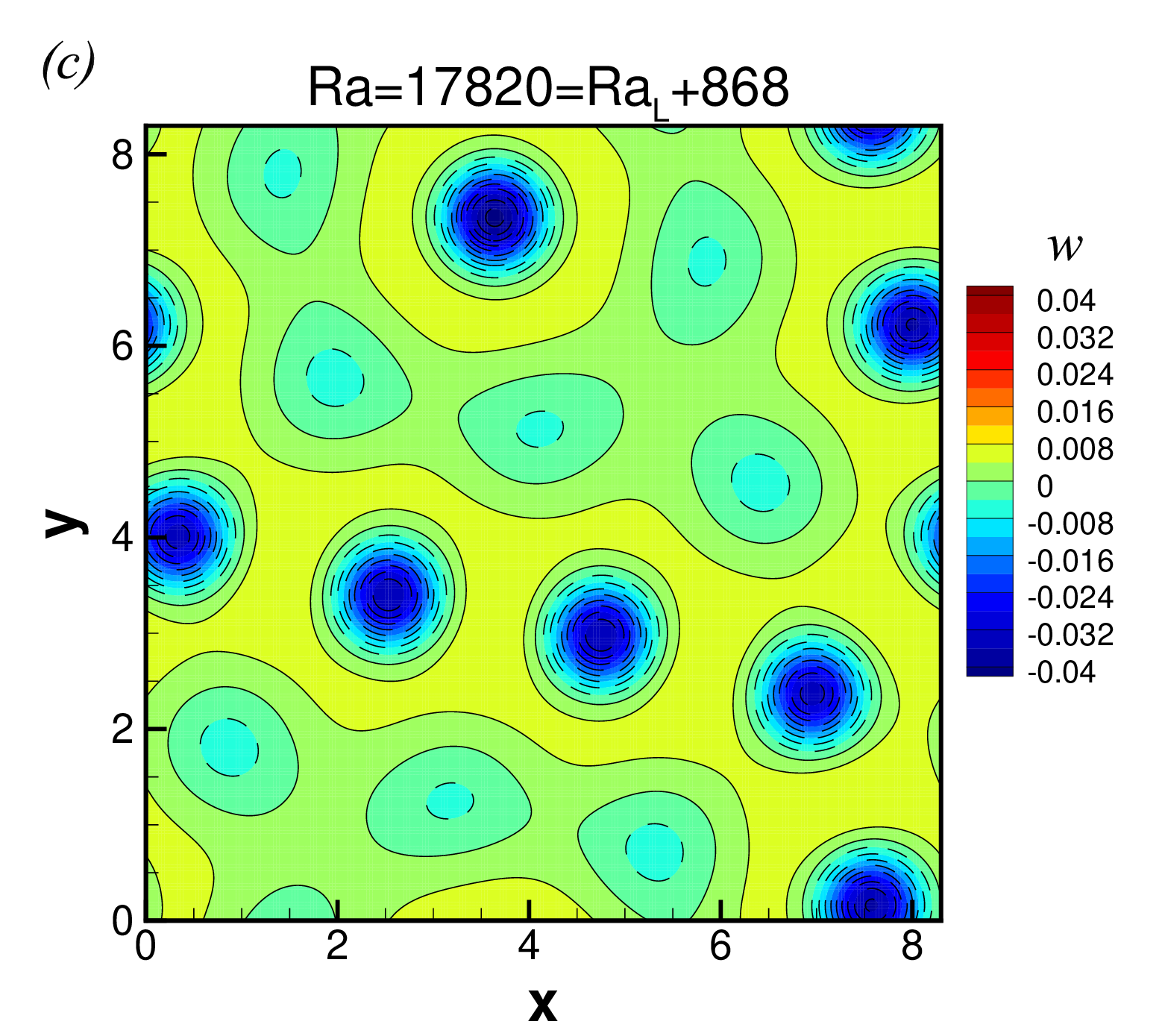}
\includegraphics[width=0.49\textwidth]{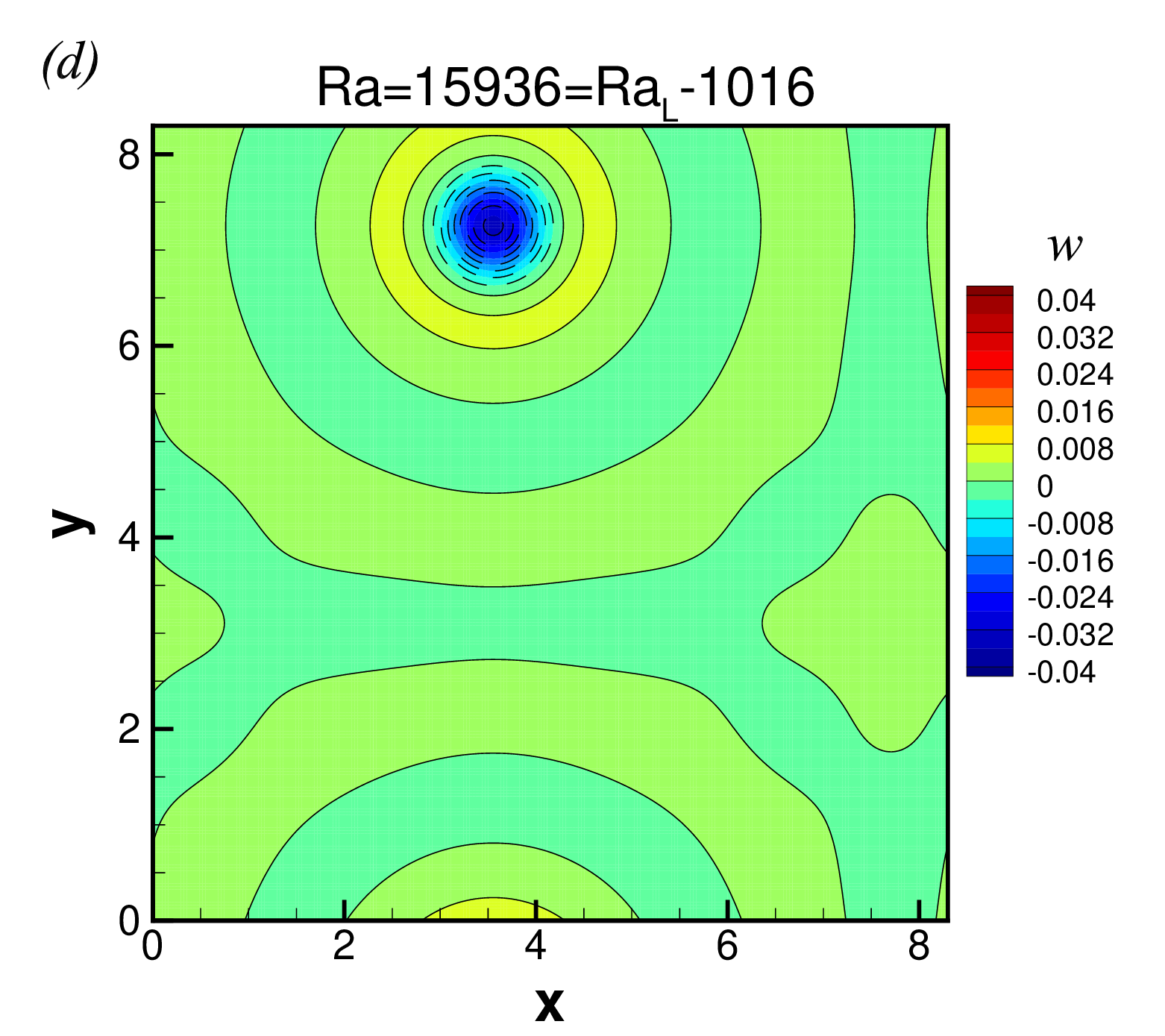}
\caption{Spatial structure of stable three-dimensional convection solutions in small (\emph{(a)} and \emph{(b)}) and large (\emph{(c)} and \emph{(d)}) domains. $\Ra=17820$ in  \emph{(a)}, \emph{(b)} and \emph{(c)}. $\Ra=15936$ in \emph{(d)}. Plots  \emph{(a)}, \emph{(c)} and \emph{(d)} show distributions of vertical velocity in the horizontal mid-plane $z=0$. Solid and dashed isolines indicate, respectively, values $w\ge 0$ and $w<0$. Plot   \emph{(b)} shows temperature and velocity vectors (every second vector in each direction is plotted) in the vertical cross-section $x=1.25$ (approximately through the center of the downflow zone visible in \emph{(a)}. }
\label{fig3}
\end{center}
\end{figure*}

{The typical spatial structure of the three-dimensional solutions is illustrated in Fig.~\ref{fig3}. We see  that the convection cells extend through the entire height of the layer, including its unstably stratified (upper) and stably stratified (lower) halves. The cells
lack reflective symmetry in the vertical plane in the sense that the maximum downflow velocity is much larger (typically about three times larger) than the maximum upflow velocity. No such asymmetry is found in the two-dimensional solutions. In the three-dimensional flows, the downflow zones are nearly circular in the horizontal cross-section (see Figs.~\ref{fig3}a,c,d). At the same time, a distorted hexagonal pattern can be discerned on the outer edges of the cells (see Fig.~\ref{fig3}c). The possibility that pre-determined (rather than random as in our simulations) spatial structure of initial perturbations may lead to other shapes of the nonlinear solutions is not explored in our study.}

In agreement with the results of the linear stability analysis, the small domain always includes one convection cell. At the parameters explored in our study (see Table \ref{tab1}) the small-domain solutions are all steady states. The behavior of the flow in the large domain is predictably more complex. At larger $\Ra$, a pattern of 4x4 cells is observed, with some cells being stronger than the others (see Fig.~\ref{fig3}c). Such flows are unsteady, exhibiting irregular evolution comprising horizontal motion, growth,  decay, and merging of individual cells. At smaller subcritical $\Ra$, the flow in the large domain becomes steady and contains a smaller number of cells. Just one cell survives at $\Ra$ approaching $\Ra_{lim}$ (see Fig.~\ref{fig3}d).

\begin{figure}
\begin{center}
\includegraphics[width=0.6\textwidth]{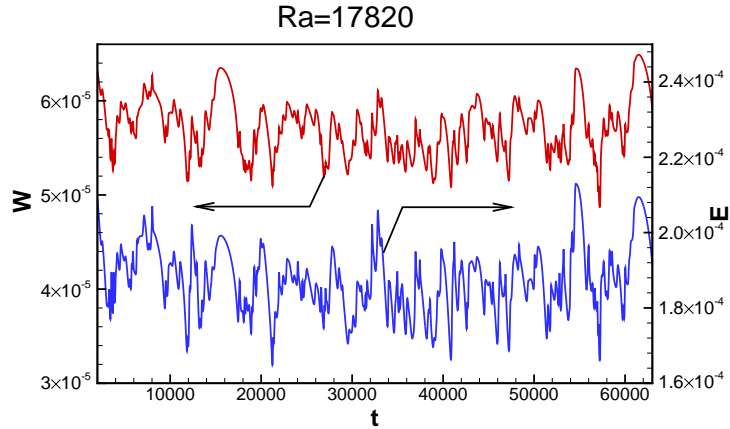}
\caption{Time signals of the integral vertical heat flux $W$ (red) and kinetic energy $E$ (blue) (see (\ref{quant})) for the flow in the large domain computed on the grid with $N_z=16$ and $N_{\bot}=32L/\lambda_L$. The part of the evolution corresponding to the fully developed flow is shown.}
\label{fig4}
\end{center}
\end{figure}

{The unsteady convection flows in the large domain at high $\Ra$ were further studied in simulations on the crude grid with $N_z=16$ and $N_{\bot}=32 L/\lambda_L$. The low resolution allowed us to explore the long-time dynamics of the flow. The results are presented in Fig.~\ref{fig4} and illustrated by the animation deposited as a supplementary material to this article. In Fig.~\ref{fig4}, we see slow (on the  scale of thousands of non-dimensional time units) irregular fluctuations  of the integral properties of the flow. The typical amplitude of the fluctuations shows no tendency to decay with time. We conclude that the time-dependency is not a transient phenomenon, but a feature of the fully developed flow presumably caused by the nonlinear interaction between the neighboring cells. This explanation is consistent with the fact that the flows in the small domain, where such an interaction is prevented by the periodic boundaries, are steady states at the same values of $\Ra$. Determining the nature and the precise location of the transition from steady to unsteady solutions in the large domain would require eigenvalue analysis and, therefore, beyond the scope of our study. As a recognition of the remaining question, the unsteady solutions are not connected to the steady solution branch in Fig.~\ref{fig1}.}

\section{Concluding remarks}
The main outcome of our work is the demonstration of the subcritical solutions of the system. The subcritical flows have the form of three-dimensional convection cells with concentrated, approximately round in the horizontal plane zones of downward flow. The cells are not strong and result in only mild increase of the Nusselt number. The difference between the flows in small and large domains is expectable and can be attributed to the various degrees of constraint imposed by the periodic boundaries. This implies that subcritical solutions at lower (but, probably, not much lower) Rayleigh numbers can be found in even larger domains.

{Our work leaves some unanswered questions possibly worthy of future exploration. It would be interesting to apply the more traditional methods of contination and eigenvalue analysis to reveal the entire steady solution branches, including their unstable portions. Other questions concern the effect of the horizontal size of the domain, the possibility of solutions with other spatial structures of the convection cells, and the transition between steady and unsteady solutions. }

The authors are thankful to David Goluskin for interesting and stimulating discussions. Financial support was provided by
the U.S. NSF (grant CBET 1435269) and by the University of Michigan - Dearborn.

%\bibliographystyle{unsrt}
%\bibliographystyle{apsrev}
%\bibliography{../../battery}
%

\end{document}